\documentclass[12pt]{article}
\usepackage{fullpage}

\usepackage{amsmath}
\usepackage{amsfonts}
\usepackage{graphicx}
\usepackage{latexsym}
\usepackage{subfigure}
\usepackage{amsthm}
\usepackage{endnotes}

\newcommand{\vls}{\vspace{3mm}}
\newcommand{\vs}{\vspace{5mm}}
\newcommand{\vbs}{\vspace{10mm}}

\newcommand{\mathlog}{\mbox{log}}

\begin{document}

\vspace{5cm}

\begin{center}

\vspace{5cm}

{\large \bf 
The perception of melodic consonance: \\
an acoustical and neurophysiological explanation \\
based on the overtone series
}

\vspace{10mm}

\hspace{1cm}\begin{tabular}{l}

Jared E. Anderson\\
\\
University of Pittsburgh\\
Department of Mathematics\\
Pittsburgh, PA, USA\\
\\

\end{tabular}

\vspace{5mm}

\end{center}

\begin{abstract}

The melodic consonance of a sequence of tones is explained using the
overtone series:  the overtones form ``flow lines'' that link the tones
melodically;  the strength of these flow lines determines the melodic
consonance.  This hypothesis admits of psychoacoustical and
neurophysiological interpretations that fit well with the place theory of
pitch perception.  The hypothesis is used to create a model for how
the auditory system judges melodic consonance, which is used to
algorithmically construct melodic sequences of tones.

\vspace{3mm}


\end{abstract}

\vspace{.5cm}

\noindent {\large \bf 1. Introduction}

\vs

\sloppy

Consonance and dissonance are a basic aspect of the perception of tones,
commonly described by words such as `pleasant/unpleasant', `smooth/rough',
`euphonious/cacophonous', or `stable/unstable'.  This is just as for other
aspects of the perception of tones:  pitch is described by `high/low';
timbre by `brassy/reedy/percussive/etc.'; loudness by `loud/soft'.  But
consonance is a trickier concept than pitch, timbre, or loudness for three
reasons:

\fussy

First, the single term {\em consonance} has been used to refer to
different perceptions.  The usual convention for distinguishing between
these is to add an adjective specifying what sort is being discussed.  
But there is not widespread agreement as to which adjectives should be
used or exactly which perceptions they are supposed to refer to, because
it is difficult to put complex perceptions into unambiguous language.  
The clearest, most musically-informed discussion of consonance and
dissonance may be found in Tenney~\cite{Tenney}.  He has identified five
distinct ``consonance/dissonance concepts'' (CDC-1, CDC-2,\dots, CDC-5)
which have been discussed throughout music history, often at cross
purposes.

Second, when hearing tones in a musical context, a listener is
simultaneously confronted with two or more of these perceptions, which may
fuse together into a single, overall perception.  For the modern listener,
familiar with Western tonal music, it is a combination of CDC-1, CDC-4,
and CDC-5 (as explained below).  But despite being difficult to discuss
unambiguously, consonance and dissonance are real perceptions, without
which music could hardly exist.  For instance, when one notices that a
wrong note has been played, this is typically because it has caused an
unexpected dissonance.

Third, unlike pitch, timbre, or loudness, consonance and dissonance are
concerned with relations between tones and require more than one tone to
have meaning.  Consequently, consonance/dissonance perceptions might be
expected to have more complicated underlying explanations (acoustical,
psychological, or neurological) than for pitch, timbre, or loudness.  
Even the seemingly straightforward perception of pitch is now believed to
result from two different underlying explanations: place theory and
temporal theory~\cite{Pierce, Terhardt}.

The most researched consonance/dissonance concepts are for simultaneously
sounding tones.  The two which are most relevant today are CDC-4 and
CDC-5.  The important distinction between them is discussed by
Bregman~\cite[502-503]{Bregman}, Krumhansl~\cite[51-55]{Krumhansl}, and 
Tenney~\cite{Tenney}, although each uses different terminology:  musical
consonance, triadic consonance, functional consonance for CDC-4;  tonal
consonance, psychoacoustic consonance, sensory consonance, timbral
consonance for CDC-5.

CDC-4 is what music theorists usually mean by consonance and dissonance;  
it is the functional harmonic consonance and dissonance that is the basis
for the theory of harmony in Western music~\cite{Piston}.  Its perception
depends on the musical context and the listener's musical background.  
Tones that are not part of the triad are heard as dissonant.  Dissonant
chords resolve to consonant ones.  One acoustical theory of CDC-4
identifies the root of a chord with a virtual pitch arising from all the
overtones of the tones of the chord~\cite{Terhardt}.

CDC-5 refers to the inherent roughness of a simultaneously sounding
collection of tones.  Its perception does not depend on the musical
context or the listener's musical background.  Helmholtz's widely accepted
acoustical theory hypothesizes that dissonance is caused by the beating of
nearby overtones of different tones~\cite{Helmholtz}.

But the oldest consonance/dissonance concept is for sequentially sounding
tones---the purely melodic consonance investigated by the ancient Greeks,
CDC-1 in Tenney.  It refers to how consonant a tone sounds relative to a
sequence of tones that precedes it---that is, how right or wrong it
sounds.  As a simple example, if a sequence begins with the three
descending notes E,D,C, then descending a semitone to a fourth note B
sounds consonant, whereas raising a semitone to a fourth note C$^\sharp$
sounds dissonant.  The probe tone experiments of Krumhansl and her
collaborators~\cite{Krumhansl} are the first experimental investigations
of melodic consonance (although she reserves the word consonance for CDC-4
and CDC-5; listeners are asked simply to rate how good or bad a tone
sounds in a particular melodic context).

\begin{figure}\begin{center}\vspace{-1.6cm}
\hspace{-10mm}
\includegraphics[height=21cm]{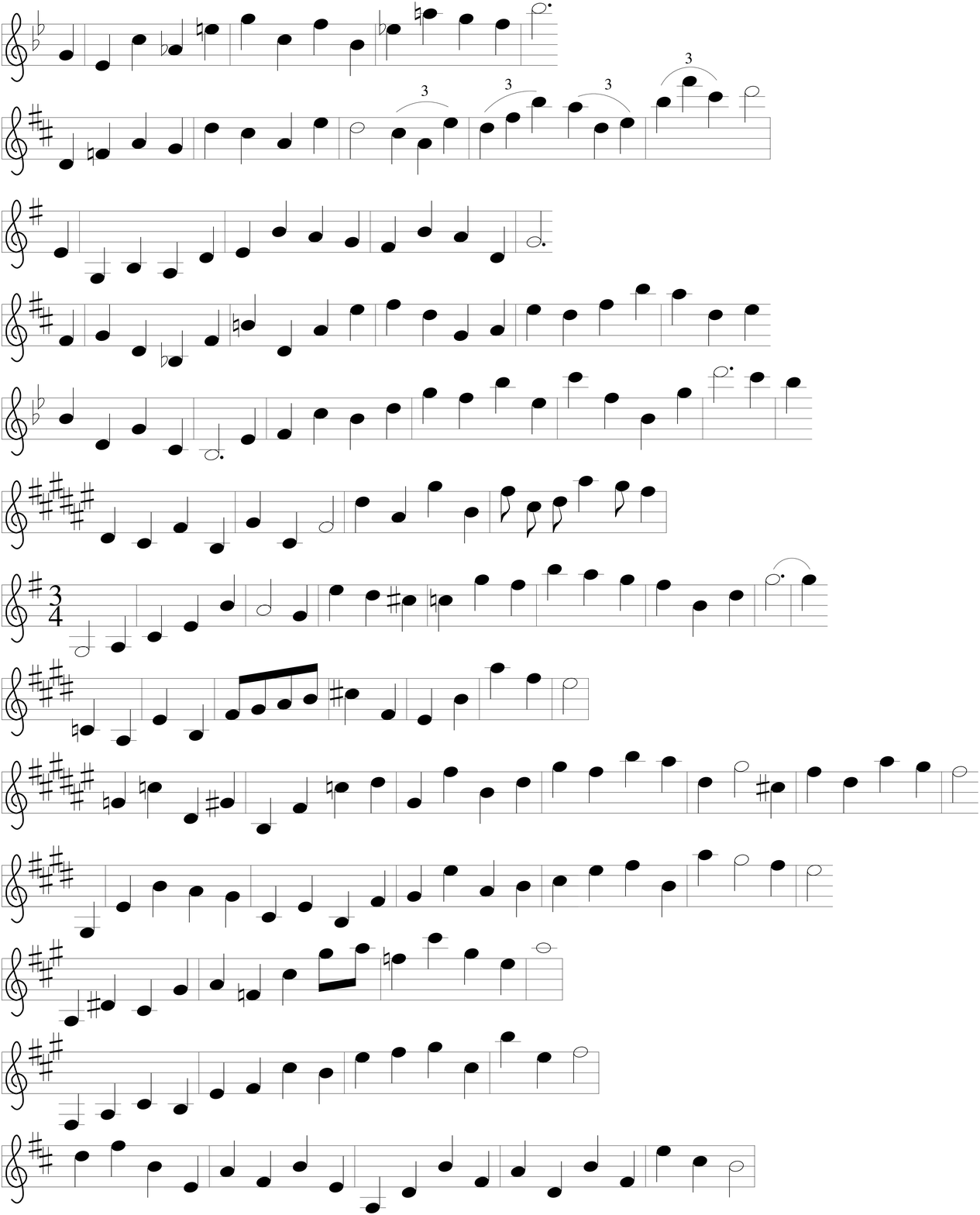}
\vspace{0cm}\caption{Initial segments of algorithmically generated sequences.} 
\label{fig:d2}
\end{center}\end{figure}

In studying melodic consonance, the simplest case (and often the only one
considered) is for a sequence of two tones;  the melodic consonance is
determined by the interval between them.  But the general case does not
reduce to this one since it is easy to find short sequences of tones that
only sound consonant in the context of a larger melody (e.g. the E$^\flat$
followed by the A in the first melody in Fig.~\ref{fig:d2}). We will
always be referring to the melodic consonance of longer sequences of
tones, or of one tone relative to a sequence that precedes it.

Melodic consonance is the fusion of (at least) two different perceptions.  
One of these is CDC-4 extended to {\em implied} harmonic progressions;  
certain tones in a melody are grouped together by nearness in time and by
rhythm, and are perceived as being part of a chord~\cite{Piston}.  People
familiar with Western tonal music find it difficult to hear a melody
without hearing these chord progressions, and most melodies are designed
to bring them out.  A tone will sound dissonant if it does not fit with
the chord the listener expects to hear.  There are many reasons why this
cannot be the only, or even the most important, perception involved in
melodic consonance: (1) the ancient Greeks studied melodic consonance
before harmony was invented; (2) beautiful, complicated melodies (perhaps
with many nonharmonic tones) may lie above a simple harmonic progression;  
(3) unmelodic sequences that follow simple harmonic progressions are
easily constructed; (4) there exist melodically consonant tonal melodies
without any implied harmony~\cite{Tovey};  (5) there exist melodically
consonant atonal melodies (for instance the opening theme to the
Schoenberg piano concerto).  Music theorists have long understood that
harmony does not suffice to explain the relations among the tones of a
melody~\cite{Hindemith, Reicha}.

This fusion of two consonance concepts is analogous to what occurs when
listening to a chord of simultaneously sounding tones in a musical
context, where CDC-4 and CDC-5 are heard together.  But there they may be
separated experimentally:  CDC-5 alone may be heard by listening to the
chord isolated from the musical context; CDC-4 alone may be heard by
playing broken chords.  Unfortunately, such separation is inherently
impossible for melodic consonance; but one may still use different words,
such as `smooth', `flowing', `lyrical', or even just `melodic', in
reference to the aspect of CDC-1 that does not depend on the listener's
functional harmonic expectations.  This is the aspect we are interested
in, and, from here on, we will refer to it simply as melodic consonance.
It is the melodic analogue of CDC-5 in that its perception should not
depend on a listener's musical background.  CDC-4 perceptions of
functional harmony, whether implied by a melody, or explicitly given by a
sequence of chords, are properly viewed as separate, even if it might be
difficult for a listener to separate them.

When hearing a tone, the ear does spectral analysis, breaking it into its
component frequencies:  the fundamental and its overtones.  In a melody,
there are usually many pairs of tones where both tones have an overtone at
(nearly) the same frequency.  (We call an interval between two tones that
have a coincident overtone a {\em musical interval}:  perfect fifth, minor
sixth, etc.)  Our motivating idea is that melodic consonance should be
determined by the pattern of overtone coincidences, and we will propose a
simple hypothesis based on this.  The idea that there should be an
acoustical explanation of melodic consonance based on memory of overtone
coincidences (at least for consecutive tones) goes back to Helmholtz, who
realized that the overtones are what give life to a melody; his book is
mostly about CDC-5, but also contains short discussions of
CDC-1~\cite[289-290,364,368]{Helmholtz}.  Our hypothesis will parallel
Helmholtz's acoustical explanation of CDC-5 in that both explanations will
be concerned with the nearness in frequency of overtones from different
tones.  Unlike Helmholtz's explanation, however, ours will interpret
consonance positively, rather than as just the absence of dissonance. A
very different acoustical explanation of melodic consonance, generalizing
CDC-5, has been explored by Sethares and McLaren~\cite{SM}.

Our hypothesis will also provide a partial explanation for the perception
of the coherence of melodies by the auditory
system~\cite[461-471]{Bregman}.  But our real interest is not in whether
or not a tone is easily incorporated into a preceding sequence, but
rather, if it is, what it sounds like in that context.

We begin by defining the concept of a flow line of overtones (section~2).  
Then we state our main hypothesis:  the melodic consonance of a sequence
of tones is determined by the strength of its flow lines (section~3).  
While this is a hypothesis of music theory, it immediately suggests
hypotheses about how the auditory system listens to and remembers a melody
(section~4).  A mathematical model based on the hypothesis is used to
algorithmically generate sequences of tones with strong flow lines, which
are heard to be melodically consonant (section~5).  Finally, we discuss
applications to composition (section~6).

\vbs

\noindent {\large \bf 2. Flow lines of overtones}

\vs

Here we use the term {\em harmonic} instead of {\em overtone} since we
want to include the fundamental as well as the upper partials.  We
restrict ourselves to the first ten or perhaps twelve harmonics since
these are the significant ones in music.

\vs

\noindent{\bf Established harmonics}

\vls

Suppose we are given a sequence of tones.  We say that a harmonic of one
of the tones is an {\em established} harmonic if it coincides with one of
the harmonics of an earlier tone.  (So the two tones are related by a
musical interval.)  How {\em well} established it is depends on:

\begin{enumerate}
\item How many times it has been previously heard.
\item The strengths of the musical intervals that establish it.
\item How many tones separate the harmonic from those that establish it.
\end{enumerate}

\noindent
The dependence in (1) is obvious: the more times previously heard, the
better.
 
In (2), the dependence is not only on the strengths of the intervals
identifying the harmonic with previous ones, but also on the strengths of
the intervals identifying these previously heard harmonics with each
other.  For example, a 9th harmonic that coincides with 4th and 5th
harmonics of previous tones is well established.  The strength of the
interval refers to the lowness of the coinciding harmonics; so fourths and
fifths are stronger than thirds and sixths for instance (in reference to
the lowest pairs of coinciding harmonics).  The unison, however, should
perhaps not be considered a very strong interval here;  so a harmonic that
is only established by previous occurrences of the same tone is not well
established.

In (3), the dependence is mainly that the fewer the number of intermediate
tones, the better.  But it is also desirable that the previous tones that
contain the given harmonic be somewhat spread out among all the previous
tones.

\vs

\noindent{\bf Flow lines}

\vls 

\begin{figure}\begin{center}\vspace{-1.6cm}
\hspace{-10mm}
\includegraphics[width=15cm]{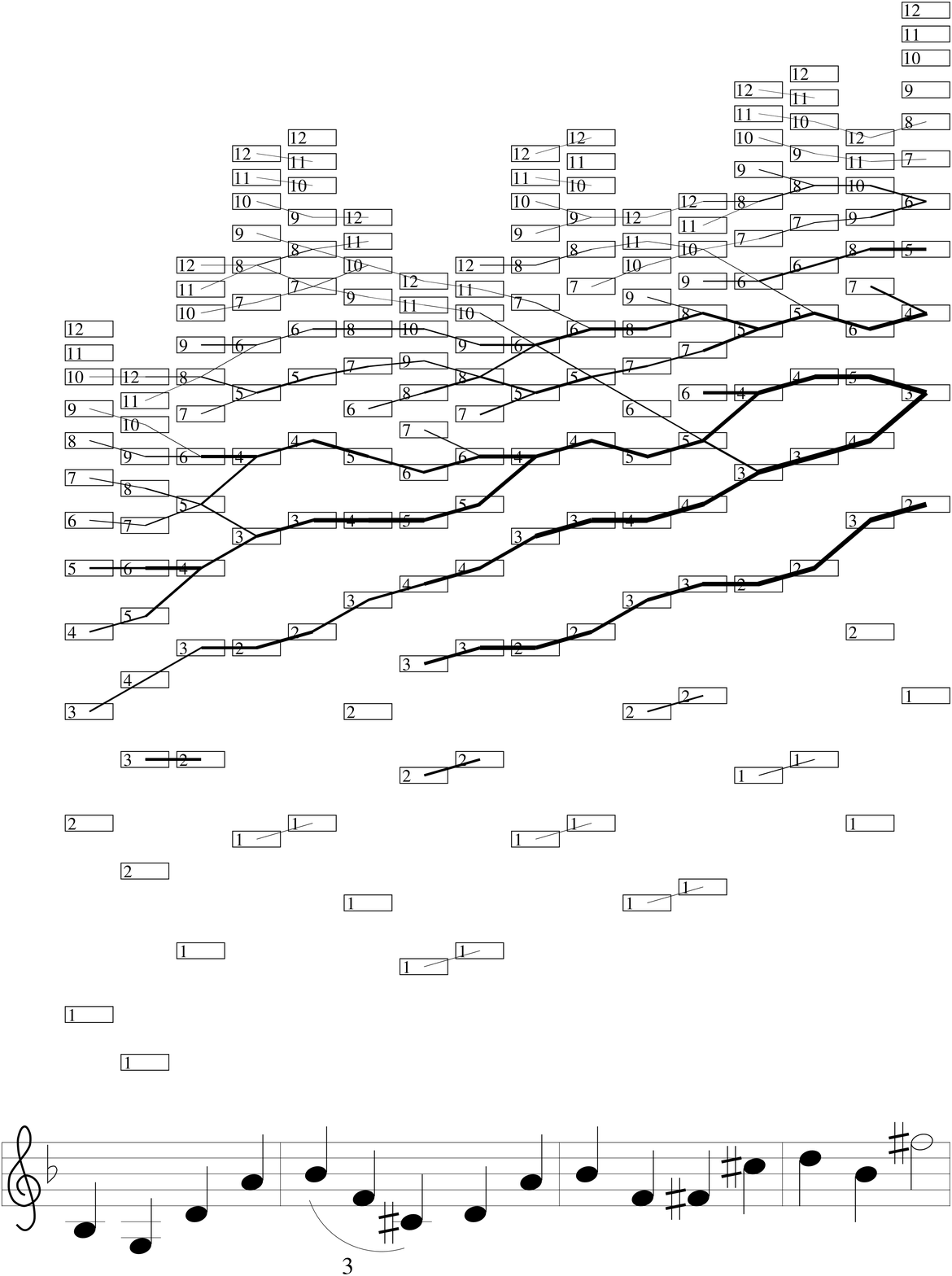}
\vspace{0cm}\caption{Algorithmically constructed flow lines.} 
\label{fig:d1}
\end{center}\end{figure}

A {\em flow line} is a sequence of harmonics from some sequence of
consecutive tones (a subsequence of our given one) where successive
harmonics are near in frequency.  `Near' usually means a whole tone or
less, although jumps of three or even four semitones may be allowed.  
(See Fig.~\ref{fig:d1} on page~\pageref{fig:d1} for a picture.) The {\em
strength} of a flow line depends on:

\begin{enumerate}
\item Its length.
\item How well established its constituent harmonics are.
\item How smooth it is.
\item How isolated it is from harmonics not on it.
\end{enumerate}

\noindent
The first two need almost no explanation:  a strong flow line is long and
a large proportion of its harmonics are well established.

In (3), `smooth' means that the flow line does not abruptly change
direction.  The smoothest flow line is always increasing, always
decreasing, or always horizontal.  If several consecutive harmonics are
all within a semitone of each other, they are regarded as essentially
horizontal.  The smoother the flow line, the stronger it is.

A consequence of (4) is that successive harmonics along a strong flow line
are near each other; almost always, the next harmonic is the nearest one
in the current direction of the flow line.  Occasionally, there may be a
nearer harmonic, within a semitone, in the opposite direction, that
interferes with the continuation of the flow line and weakens it at that
point.  For example, if the 4th harmonic of C is followed by the 6th
harmonic of the G below it, along an increasing flow line, there is a
weakness due to the nearby 5th harmonic of the G.

\vbs

\noindent {\large \bf 3. Hypothesis}

\vs

\noindent
{\bf Hypothesis:}  
{\it The melodic consonance of a tone that continues a sequence of tones
is determined by how well its harmonics add to the strengths of the flow
lines already established.}

\vs

Exactly what `how well' means is deliberately imprecise:  Just how do
harmonics fit together to form flow lines?  Are fewer strong flow lines
preferable to many weaker ones?  Is the positive effect of a harmonic
being on a flow line comparable to the negative effect of a harmonic not
being on any flow line? May flow lines cross, fork, or join together?  Is
there a preference for convergent, divergent, parallel, or widely
separated flow lines?  The model in Section~5 will give a computational
meaning to `how well' and will effectively give one possible set of
answers to these questions.  But without committing to any such precise
interpretation, the hypothesis may be clarified by discussing how the
factors that determine the strength of a flow line are supposed to affect
melodic consonance and dissonance.

So suppose we are given a sequence that ends with tone~$t$, and we add on
another tone~$t'$.  Some of the harmonics of $t$ will lie on strong flow
lines, and each flow line will have a direction at $t$:  increasing,
decreasing, or horizontal.  Harmonics of $t'$ that continue these flow
lines contribute to its melodic consonance, more so if well established.  
(`Continue' means the harmonic should be near the previous one and in the
right direction.)  The following contribute to its melodic dissonance:  
(1) a strong flow line is not continued by any harmonic of $t'$;  (2) a
well established harmonic of $t'$ does not continue any flow line;  (3) a
harmonic of $t'$ that is nearby but in the wrong direction interferes with
the continuation of the flow line through another harmonic. Naturally,
these possibilities may occur at the same time in varying degrees when a
tone sounds.

The interest in the hypothesis is in the detailed information it provides
about the sequence of tones in a melody.  But we can immediately see that
it is consistent with some standard properties of melodies: (1)~The
preference for motion by small intervals is because this is the easiest
way for each overtone to continue along a flow line; huge skips result in
broken flow lines.  (2)~Melodies use notes from scales so that there are
many well established harmonics. (3)~Small melodic variations in pitch
(away from equal temperament, as when a melody is played on a violin) may
be so that coincident harmonics on important flow lines coincide exactly.

\vbs

\noindent {\large \bf 4. Acoustical and neurophysiological interpretations}

\vs

The hypothesis that flow lines are important in music belongs purely to
music theory since it makes no reference to the auditory system.  Still,
it effectively provides a simple model for the perception of melodic
consonance and, indeed, for the perception and memory of melody:  At any
instant, when listening to a melody, the auditory system is tracking a
number of flow lines and paying attention to them according to their
strength.  At the same time, it remembers the frequencies that have been
heard as harmonics of previous tones in the sequence, paying attention to
them according to how well they are established.  The ease with which the
auditory system expects or remembers the next tone depends on how well its
harmonics continue strong flow lines with well established harmonics;  
this degree of ease is perceived as the melodic consonance of the tone.

Thus, even in monophonic music, the mind is following a kind of
counterpoint among the individual harmonics, and the stronger the flow
lines, the easier this is.  Just as one is not aware of hearing individual
overtones, one is not aware of hearing the flow lines that connect them.
This model may be interpreted at any level of the auditory system, from
the basilar membrane up through the auditory cortex, since tonotopic
organization is preserved throughout~\cite{Ribaupierre}.

At the lowest level, a complex tone produces excitations in a sequence
along the basilar membrane, at positions corresponding (logarithmically)
to the frequencies of the harmonics:  $f_0, 2f_0, 3f_0, 4f_0, ...$.  So a
flow line may be interpreted as a moving sequence of excitations.  Since
the region of the basilar membrane excited by each harmonic has a finite
spatial extension of a few semitones, there is overlap between the regions
excited by consecutive harmonics along a flow line.  Therefore, although
the centres of these regions move in discrete steps, the flow line itself
may be viewed as a region of excitation moving somewhat continuously along
the basilar membrane.

\begin{figure}
\begin{center}
\includegraphics[width=8cm]{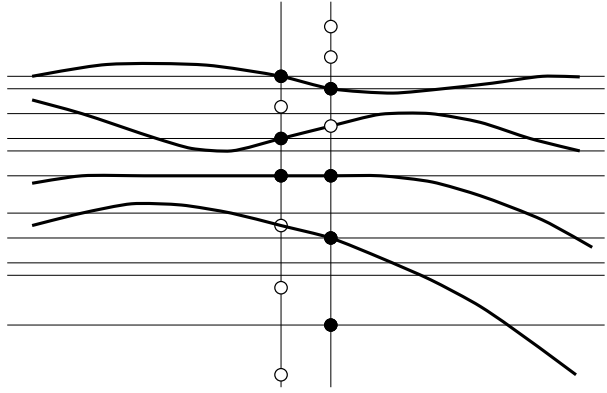}
\caption{Flow lines.} 
\label{fig:flow}
\end{center}
\end{figure}

This interpretation fits well with the place theory of pitch
perception~\cite{Terhardt}, according to which the auditory system, being
familiar with the logarithmic pattern of the harmonics, uses spatial
pattern recognition to determine the pitch of a complex tone.  Not all of
the harmonics are needed, two or three being enough to determine the
pitch;  the fundamental may be entirely absent.  So if the auditory system
is following two or more flow lines and expects a tone at a particular
time, it may determine its expected pitch based on the separation of the
lines.  This may be seen graphically in Fig.~\ref{fig:flow}, where the
horizontal lines, representing established harmonics, intersect the curved
lines, representing the flow lines.  (The horizontal axis is time and the
vertical axis is either log frequency or position along the basilar
membrane.  The vertical lines represent two consecutive tones, with
harmonics 2 through 8 indicated by circles; the established ones are
filled in.)

At the highest level of the auditory system, in the auditory cortex, a
harmonic corresponds to a localized region of neuron activity, at a
position corresponding to the frequency of the harmonic.  A flow line is
interpreted as a moving region of neuron activity; nearby neurons
correspond to nearby harmonics.  The auditory system is not well enough
understood to know what exactly would be the proper location for this
interpretation: the primary auditory cortex, or one of the higher belt or
parabelt regions~\cite{KHT, ZEM}.

But it is crucial to this interpretation that a harmonic complex tone be
encoded in the primary auditory cortex by its sequence of harmonics,
rather than by its pitch.  The most recent research indicates that this is
the case~\cite{FRAS}.  If so, it would be surprising if the brain did not
use this information temporally when listening to music.  (Most music
theory tacitly assumes exactly this---that music may be understood solely
in terms of the pitches, with the temporal pattern of the overtone
frequencies completely ignored.)

Shepard's famous auditory illusion of an endlessly ascending chromatic
scale~\cite{Shepard} provides evidence that the auditory system does pay
attention to flow lines of overtones when listening to a sequence of
tones.  Shepard used twelve tones with harmonics only at the octave
multiples and amplitudes determined by a fixed broad envelope;  after
increasing each harmonic by a semitone, twelve times, one has returned to
the initial tone since the amplitude envelope has not changed.  If this
sequence is played repeatedly, the listener perceives tones that seem to
ascend forever.  It seems inescapable that the auditory system must be
tracking the harmonics themselves, which form a series of ever-increasing
flow lines.  Exactly how the auditory system's pitch processing mechanism
is interpreting this sequence is, for us, beside the point.

An obvious objection to these psychoacoustical and neurophysiological
interpretations of the hypothesis is that they might seem to suggest that
melodic consonance is highly dependent on the timbre of the tones.  In
particular, what if some harmonics are absent, as for the clarinet?  And a
melody of pure tones is certainly not wholly unmelodic---Helmholtz's
remarks~\cite[290]{Helmholtz} notwithstanding.  This objection is easily
overcome by realizing that the auditory system might well imagine a
harmonic even if it is absent from the stimulus.  This would require a
feedback mechanism from the pitch processor whereby, having determined the
pitch based on the harmonics present, the auditory system determines the
positions of any missing harmonics by filling in the familiar pattern.  
That the auditory system may imagine such information is clear to any
musician who can read a score;  for then {\em every} harmonic is absent.

Of course, even if the hypothesis and these interpretations of it are
correct, flow lines of overtones cannot be the only way in which the
auditory system perceives and remembers a melody.  It is self-evident that
rhythm, implied harmony, melodic contour, and motives are all perceived
consciously.

\vbs

\noindent {\large \bf 5. A mathematical model and algorithmically
generated melodies}

\vs

The hypothesis is used to construct a mathematical model for how the
auditory system tracks flow lines and determines melodic consonance.  We
do this for several reasons:

First, the model is effectively a specific version of the hypothesis,
eliminating the imprecision in its statement and changing qualitative
predictions to quantitative ones.  Naturally, a general qualitative
hypothesis is more likely to be correct than the details of any particular
implementation of it in a mathematical model.  Our implementation is
not the only one possible.

Second, the model provides indirect evidence supporting the hypothesis, in
the spirit of Krumhansl's probe tone experiments~\cite{Krumhansl}.  In
brief, it allows for the algorithmic construction of sequences of tones
that have strong flow lines.  This is done recursively, choosing at each
step a tone that the model predicts will sound melodically consonant
relative to the preceding sequence; so each successive tone may be viewed
as a predicted result of probe tone experiment at that point.  That
sequences that do sound melodically consonant may be produced in this way
is evidence in support of the hypothesis. Of course this evidence is not
conclusive since it is possible that the melodic consonance is due not to
the flow lines, but to other properties of the sequences that are a
by-product of the construction.  One could argue, for instance, that the
melodic consonance is due only to there being many well established
harmonics on the constructed tones---a much weaker hypothesis.  Glaringly
wrong notes are indeed ones with no established harmonics.

Third, this provides a new method for the algorithmic composition of
melodic sequences of tones.  This method appears to be unique in that:
\begin{enumerate} 
\item Absolutely no musical information is put into the model:  no data
from existing melodies, no information about which intervals are to be
allowed or with what probability they should be chosen, and no information
about tonality, implied harmony, or restrictions of the tones to a fixed
diatonic scale.  (It is true that the tones are restricted to an equal
tempered chromatic scale, rather than allowing arbitrary pitches; but
this is really just a matter of convenience and it should be possible to
obtain similar results even if initially allowing for completely
arbitrary pitches.)
\item Nevertheless, reasonable melodies result.  They generally share
these features with melodies composed by people:  (i)~They are tonal,
selecting most tones from some (usually major) diatonic scale; the tonic
and dominant occur frequently.  (ii)~Nonharmonic tones are occasionally
selected (but melodically consonant ones). (iii)~The intervals between
consecutive tones are usually relatively small and there are often short
scale-like passages. (iv)~They sound melodically consonant, increasingly
so with repeated listening.  
\end{enumerate}

Fourth, this demonstrates that the hypothesis should be of practical
interest to music theory and composition, and not just to psychoacoustics
and neurobiology.  It predicts details of melodic structure that are
inaccessible to conventional music theory.  Although it hasn't been
presented in this way, this research began purely as music theory, by
analysing the pattern of overtone coincidences in hundreds of existing
melodies.

\vbs

\noindent{\bf The Model}

\vs

\noindent
\underline{Setup}

\vls

Suppose we are given a sequence of tones $t_1, t_2, t_3, ...$ in equal
temperament:  each $t_i$ is an integer specifying the number of semitones
above or below some fixed reference tone.  In these units, the pitch of
the $k$th harmonic of the $i$th tone is then $p_{i,k}=t_i+12\log_2k$.  
Fix $\epsilon=0.2$ semitones, which we call the {\em equal temperament
error}.  Although $p_{i,k}$ is not an integer unless $k$ is a power of
$2$, we say that $p_{i,k}$ is an {\em integral pitch} if there is an
integer $p$ such that $|p-p_{i,k}|<\epsilon$.  Then, for $1\leq k \leq 12$
(which are the harmonics we will consider), $p_{i,k}$ is an integral pitch
if and only if $k\neq 7,11$;  in this case the integer $p$ is denoted
$\mbox{Round}(p_{i,k})$.  We say that the $k$th harmonic of the $i$th tone
{\em coincides} with the $l$th harmonic of the $j$th tone if $p_{i,k}$ and
$p_{j,l}$ are integral pitches for which
$\mbox{Round}(p_{i,k})=\mbox{Round}(p_{j,l})$.

For each $1\leq k \leq 12$, define a constant $h_k$ between $0$ and $1$
that specifies the importance of the $k$th harmonic to the flow lines.  
These values are not the amplitudes of the harmonics, although they may be
thought of as partially determined by them.  For the data below
$(h_1,h_2,... ,h_{12}) = (0.3, 0.8, 0.9, 0.8, 0.7, 0.6, 0.3, 0.5, 0.4,
0.3, 0.05, 0.05)$.

\vs

\noindent
\underline{Established harmonics}  

\vls

Now we define a value $E_{i,k}$ between $0$ and $1$ that is a measure of
how well established the $k$th harmonic of the $i$th tone is.  The
formulas are not particularly important, and others would work as well.  
All that's necessary is that $E_{i,k}$ be close to $1$ if the harmonic is
qualitatively well established, and close to $0$ if it is not.  We set
$E_{i,k}=\mbox{Min}\{1, \sum \phi(k_1,k_2) e^{-\kappa (i - i_1 - 1)} \}$
where the sum is over all $(i_1,k_1)$ and $(i_2,k_2)$ with $1\leq i_1 <
i_2 \leq i$ for which $p_{i_1,k_1}$, $p_{i_2,k_2}$ and $p_{i,k}$ coincide.  
Here the decay constant $\kappa = 0.15$, and $\phi(k_1,k_2)=h_{k_1}\cdot
h_{k_2}$ if $k_1\neq k_2$;  $\phi(k_1,k_2)= \mbox{Min}\{0.1,h_{k_1}\cdot
h_{k_2}\}$ if $k_1=k_2$. (The idea of the exponential factor $e^{-\kappa
(i - i_1 - 1)}$ is that it is the product of two factors:  (1) the decay
$e^{-\kappa (i_2 - i_1 - 1)}$ owing to the separation between the two
coincident harmonics $p_{i_1,k_1}$, $p_{i_2,k_2}$;  (2) the decay
$e^{-\kappa (i - i_2)}$ owing to the more recent of these being heard
$i-i_2$ tones before.  It is $1$ only when $i_1=i-1$ and $i_2=i$.)

\vs

\noindent
\underline{Flow line construction}  

\vls

The flow lines are constructed recursively.  Each flow line is viewed as a
sequence $(a(n), a(n+1), a(n+2), ..., a(n+l-1))$ where $n$ is the index of
the first harmonic, $l$ is the length, and each $a(i)\in\{1,2,...,12\}$.  
Every harmonic of every tone will be on at least one flow line (although
it may have length~$1$).  With the first tone, $12$ flow lines are begun,
one for each harmonic.  Now, suppose the flow lines have been constructed
for the first $i$ tones, and we add tone $i+1$.  For each flow line $a$
that is still active (i.e., that contains a harmonic of the $i$th tone) we
either add another harmonic $a(i+1)$ or end the flow line (so that $a(i)$
is the last one)  according to the following rules.

Let $\overline{k}$ be the smallest integer such that $p_{i+1,\overline{k}}
\geq p_{i,a(i)} - \epsilon$ and $\underline{k}$ the largest integer such
that $p_{i+1,\underline{k}} \leq p_{i,a(i)} + \epsilon$, {\em if} such
integers exist in $\{1,2,...,12\}$.  If both exist, it is easy to see that
either $\overline{k}=\underline{k}$ or $\overline{k}=\underline{k}+1$. Let
$M = 3 + \epsilon$ be the maximum allowed jump (in semitones) between
consecutive harmonics of a flow line, and let $M^\prime = 2 + 2\epsilon$
be the maximum allowed change in direction in a flow line.  If the flow
line contains an $a(i-1)$---i.e. $a(i)$ did not begin it---then let
$\Delta = p_{i,a(i)} - p_{i-1,a(i-1)}$; otherwise let $\Delta=0$; so
$\Delta$ is the previous jump in the flow line.  If $\overline{k}$ exists,
let $\overline{\Delta} = p_{i+1,\overline{k}} - p_{i,a(i)}$;  if
$\underline{k}$ exists, let $\underline{\Delta} = p_{i+1,\underline{k}} -
p_{i,a(i)}$.  Let us say that {\em `up' is allowed} if $\overline{k}$
exists, $|\overline{\Delta}| \leq M$, and $|\overline{\Delta} - \Delta|
\leq M^\prime$;  similarly say {\em `down' is allowed} if $\underline{k}$
exists, $|\underline{\Delta}| \leq M$, and $|\underline{\Delta} - \Delta|
\leq M^\prime$.  If neither is allowed, the flow line stops at $a(i)$.  
If `up' is allowed but not `down', the flow line continues with $a(i+1) =
\overline{k}$.  If `down' is allowed but not `up', the flow line continues
with $a(i+1) = \underline{k}$.  If both are allowed, let
$p=p_{i,a(i)}+\alpha\Delta$ where $\alpha=0.4$ is a positive constant.  
(This constant is important in determining smoothness;  a small value of
$\alpha$ makes it more desirable for a steep flow line to gradually level
off.  I believe $\alpha > 0.5$ gives incorrect flow lines in some
instances.)  This $p$ should be thought of as the pitch that would provide
the smoothest continuation.  So we compare $|p-p_{i+1,\overline{k}}|$ with
$|p-p_{i+1,\underline{k}}|$;  if the first is smaller, we continue with
$a(i+1)=\overline{k}$, otherwise with $a(i+1)=\underline{k}$.  This is not
quite right, since if the difference is too close to call ($\big|
|p-p_{i+1,\overline{k}}| - |p-p_{i+1,\underline{k}}| \big| \leq 0.8$)  we
instead continue with whichever harmonic is better established, by
comparing $E_{i+1,\overline{k}}$ and $E_{i+1,\underline{k}}$.  (Of course
if $\overline{k} = \underline{k}$ then there are no comparisons to make.)

Having continued all active flow lines, we let any harmonic of $t_{i+1}$
that did not continue a flow line, begin a new one.

As stated, it is entirely possible for two flow lines to converge and
agree after some point.  Clearly if they agree for two consecutive
harmonics, then they will agree thereafter.  So if $a$ and $b$ are flow
lines, and $a(i)=b(i)$ and $a(i+1)=b(i+1)$, then we stop one of them with
last harmonic $a(i)=b(i)$, although we keep track of the fact that it
joined with the other one.

\vs

\noindent
\underline{Strength of a flow line}  

\vls

Each harmonic $a(i)$ of each flow line $a$ will be assigned a strength
$S(a,i)$ between $0$ and $1$ that measures the strength of the flow line
at that point; $S(a,i)$ will be a linear combination of the $h_{a(j)}$ for
$j\leq i$.  If $a=(a(n),a(n+1),...,a(n+l-1))$, define $S(a,n) = \mu
h_{a(n)}$ where $\mu=0.25$ is a constant that determines the initial
strength of flow lines.  We recursively define $S(a,i)$ for $i>n$:  given
$S(a,i)$, we let $S(a,i+1)= k S(a,i) + (1-k) (ABC) h_{a(i+1)}$ where
$k=0.6$ is a constant that specifies the relative importance of the flow
line's prior strength to how well the next harmonic continues it.  
$A,B,C$ are values between $0$ and $1$ determined as follows.

\sloppy

$A$~is a measure of how smoothly harmonic $a(i+1)$ continues the flow
line:  $A=k_A \exp(-s|p_{i+1,a(i+1)} - p|^2) + (1-k_A)$ where the decay
constant $s=0.35$, $p=p_{i,a(i)}+\alpha\Delta$ is defined as before, and
$k_A=0.5$. $B$~is a measure of how well established the harmonic is:  
$B=k_B E_{i+1,a(i+1)} + (1-k_B)$ where $k_B=0.6$. $C$~is a factor that
diminishes the strength of the flow line if another harmonic interferes
with its continuation:  If tone $t_{i+1}$ contains a harmonic other than
$a(i+1)$ that is within $1+\epsilon$ semitones of $p_{i,a(i)}$ (it would
be $a(i+1)+1$ or $a(i+1)-1$) then $C=1-k_C=0.5$; otherwise $C=1.0$.  The
constants $k_A, k_B, k_C$ being less than $1$ ensures that the failure of
a harmonic to continue a flow line well in some way does not preclude it
from still contributing to its strength.

\fussy

If flow lines join together, the strength of the continuing flow line is
the greatest of the strengths of the incoming flow lines.

\vs

\noindent
\underline{How well a tone continues the flow}

\vls 

For each tone $t_{i+1}$ we define a quantity $\Phi_{i+1}$ that is a
measure of how well it continues all of the flow lines for tones $t_1,
t_2,..., t_i$.  Let $S=\sum_a S(a,i)$ where the sum is over all active
flow lines $a$.  Then, for each such $a$, let $s_a=S(a,i)/S$ be the
relative strength of the flow line.  Let $\Phi_{i+1} = \sum_a s_a ABCD$
where $A,B,C$ depend on $a$ as discussed above, and $D=k_D h_{a(i+1)}+
(1-k_D)$ with $k_D = 0.5$.  Clearly $\Phi_{i+1}$ is between $0$ and $1$.  
(One might worry that one of the $s_a$ might be close to $1$, and dominate
all others.  In practice, with the above values for constants, this does
not happen; if it did, one would need to impose a maximum allowable value
for $s_a$.)

According to the hypothesis, $\Phi_{i+1}$ is a measure of the melodic
consonance of tone $t_{i+1}$.  Actually, we have not accounted for the
part of the hypothesis stating that well established harmonics that do not
continue any flow line are a distraction and cause melodic dissonance.  
One might subtract a constant times $B h_k$ from $\Phi_{i+1}$ for each
such harmonic $k$, but, for simplicity, we have omitted this.  In any
case, with the above setup, such harmonics may begin flow lines, and do
not contribute to $\Phi_{i+1}$.  One might also subtract something for
flow lines $a$ that stop at tone $t_i$, although the model does account
for this by the absence of a contribution to $\Phi_{i+1}$.

One should not read too much into these values $\Phi_i$.  They provide
only a crude way of comparing the melodic consonance of different tones
that might follow a given sequence.  They are not meaningful in comparing
melodic consonance in different sequences or at different positions of one
sequence.

Also, in even defining $\Phi_i$, we have implicitly assumed that melodic
consonance may be ordered, i.e., that it always makes sense to say that
one tone sounds more melodically consonant than another in a particular
context.  The hypothesis, however, allows that a tone might sound
melodically consonant in different ways according to {\em how} its
overtones strengthen the flow lines.  Nevertheless, this assumption is a
useful simplification.

\vbs

\noindent{\bf Algorithmically generated sequences}

\vs

We begin with two selected tones $t_1, t_2$ and recursively generate a
sequence: if we know the first $i$ tones, we can choose as tone $t_{i+1}$
that one for which $\Phi_{i+1}$ is maximum.  We allow $t_{i+1}$ to range
between $-100$ and $+100$, with $0$ corresponding to an A.  To ensure some
melodic interest, we do not allow the same pitch or any octave equivalent
to repeat in any four consecutive tones.  So as not to always generate the
same sequences, we randomly select each tone from among those for which
$\Phi_{i+1}$ is within 90\% of the maximum possible value; typically there
are $1$ or $2$ choices, occasionally $3$ or more.

The algorithm was programmed in C on a UNIX operating system.  The
pseudorandom number generator was seeded with $0$ for the data below.  
Thirty sequences were generated sequentially, with the first tone $t_1$
chosen randomly between $-6$ and $5$, and the difference $t_2-t_1$ cycling
through different intervals.  Initial segments of half of the $28$
admissible sequences (the two with \mbox{$t_1=t_2$} being discarded)  are
in Figs.~\ref{fig:d1} and~\ref{fig:d2}.  These are the $14$ that seemed to
make the best melodies, although the others are comparable.  They were
ended at the point that sounded best, often just before the sequence rose
to a higher range (there being a clear, overall tendency for pitch to
drift upwards).  Since the output is invariably tonal, they are notated
with the key signature that seems to fit best.  The rhythm is the simplest
that sounded reasonable.

In Fig.~\ref{fig:d1}, the vertical axis is log frequency or position along
the basilar membrane.  Each flow line $a$ is drawn from $p_{i,a(i)}$ to
$p_{i+1,a(i+1)}$ with thickness proportional to $S(a,i+1)$.  Those
$p_{i,k}$ with integral pitch are drawn at height $\mbox{Round}(p_{i,k})$
so that coincident harmonics line up exactly in the picture.  (A
straightedge may help in looking at how well established each harmonic
is.)

\begin{figure}\begin{center}\vspace{-1.6cm}
\hspace{-10mm}
\includegraphics[height=22cm]{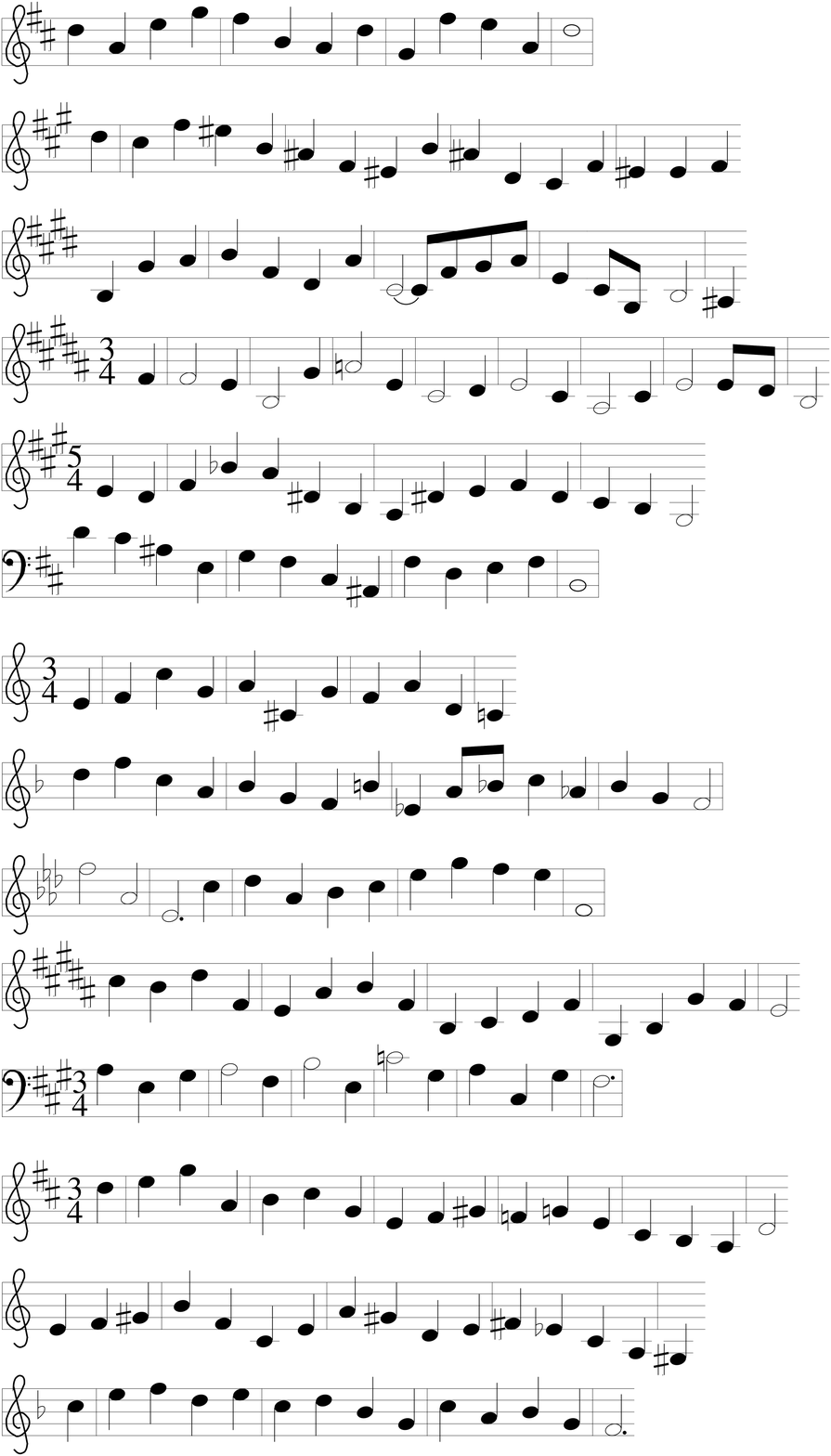}
\vspace{0cm}\caption{Melodies analytically generated by hand.} 
\label{fig:d3}
\end{center}\end{figure}

\vbs

\noindent {\large \bf 6. Applications to composition}

\vs

\noindent{\bf Algorithmic composition}

\vls

We saw in the previous section that melodically consonant sequences of
tones may be generated algorithmically.  But this algorithm, as it stands,
would not be useful to composers since the resulting melodies are only
mediocre. One reason is that although melodic consonance is only one
aspect of melody, it was the only aspect considered in constructing the
sequences.  No consideration was given to melodic contour, rhythm,
or---especially---implied harmony.

\sloppy

The theory of harmony is concerned with relations between roots of
different groups of tones, and, for implied harmony, the grouping of tones
is largely achieved through rhythm~\cite{Piston}. No doubt the ear chose
rhythms for the sequences so that they made sense harmonically, i.e., so
that chord progressions are clearly heard.  The success of a melody
depends largely on how well the harmony is brought out. While flow lines
have little to do with the theory of harmony, they are inextricably linked
to it in practice simply because both the flow and the harmony arise at
the same time from the same tones.

\fussy

Since melodic contour is completely ignored, the generated sequences
perhaps use more large skips than is desirable.  Sevenths, for instance,
are sometimes judged no more melodically dissonant than the corresponding
second an octave away.  In other examples, large skips of more than an
octave occasionally occur.  This may indicate that melodic contour makes a
contribution to melodic consonance, independent of the flow lines.

Also, many of the generated sequences sound rather similar, for the
algorithm makes uninteresting choices.  And even if one had a perfect
measure of melodic consonance, ``more melodically consonant'' certainly
does not mean ``better''.  (This is, however, much closer to being true
than for CDC-4, where functional harmonic dissonances are essential. But
in most melodies, the tones ought to flow together melodically, at least
in stretches;  although it may be useful to sacrifice some melodic
consonance for melodic interest, melodic dissonance in itself is rarely
appreciated.)

Finally, even if the hypothesis is correct, its implementation in the
algorithm is undoubtedly wrong in many respects; much more work should be
done in trying to find the algorithm based on the hypothesis that achieves
the best possible results.

\vs

\noindent{\bf Constructing melodies analytically, by hand}

\vls

The hypothesis is more useful in constructing melodies by hand.  One can
make a kind of slide-rule/abacus with twenty or more vertical strips that
may be slid up and down.  Each should be marked with the first ten
harmonics:  divide it into 41 equal segments and colour in numbers
1,13,20,25,29,32,34.7,37,39,41 (the $n$th one being nearly
$1+12\thinspace{\mathlog}_2 n$).  These were made out of thin cardboard
and allowed to slide freely under a transparency attached at both ends to
a thick piece of cardboard.  A piano keyboard background was drawn on the
cardboard to allow for easy translation to musical notation.  Presumably
something more durable can be made.

The hypothesis was formulated by doing listening experiments with this
slide-rule and using it to try to construct melodically consonant
sequences of tones by hand (before listening to them or imagining the
sound).  Such experiments are the only way to develop a detailed
understanding of the hypothesis, and were obviously prerequisites to
writing any algorithm.

Using the slide-rule to construct sequences of tones with interesting flow
lines is a task ideally suited to the human visual system, and becomes
easy with practice.  There are two obvious advantages to this over using
the algorithm:  First, it is easy to take the melodic contour into
consideration.  Second, it is easy to sometimes select less melodically
consonant but more interesting tones, making use of less common overtone
coincidences.  And with a little practice, one is generally able to
visually identify whether or not a tone is a scale tone or a nonharmonic
tone, the latter having fewer established harmonics.  The tonic and
dominant may often be visually identified as the tones with by far the
most well established harmonics.

Examples of some of the better melodies created in this way are in
Fig.~\ref{fig:d3}.  (Often, however, the ear was used, either in
discarding the last several tones of a sequence, or in adding one
additional tone to the end.)  Perhaps some composers (of tonal and atonal
music) will find this idea a useful addition to the theory of harmony,
twelve-tone composition and other analytical methods.

\vs

\noindent{\bf Xenharmonic music}

\vls 

Another potential application is to xenharmonic music, as described by
Sethares for instance~\cite{Sethares}.  The idea is to abandon the
harmonic spectrum and use tones with nonharmonic spectra.  One can use
natural nonharmonic tones or modify them electronically to produce tones
with a desired spectrum.  As Sethares explains, the spectra of the tones
determines the scale that should be used, and vice versa.  (The drawback
of xenharmonic music is that it may be difficult to appreciate until the
auditory system learns to recognize the spectrum.)  If flow lines
determine melodic consonance here, this suggests what sort of spectra are
desirable: there ought to be many intervals for which one or more pairs of
overtones coincide and also for which overtones of one tone are near
overtones of the other.  In terms of Sethares's dissonance curves, the
first of these means that the minima should be {\em sharp} ones.  The
second involves some compromise since nearby overtones are causes of both
harmonic dissonance and melodic consonance.

\vbs

\noindent {\large \bf Acknowledgements}

\vls

I owe a lot to my music teachers.  I am also grateful to four people with
whom I have discussed this work; their comments have lead to significant
improvements in the exposition.  I am not mentioning any names since I
have not asked for their permission to do so and I do not wish to bother
them about it now.

\vs

\noindent {\large \bf Remark}

\vls 

This research was done between 2000 and 2002.  I am now (2004) no longer
actively researching musical acoustics, although I may come back to it.  
Still, feel free to e-mail me about this.  As a postdoc, my academic
e-mail address keeps changing; but the address {\tt mjaredm@yahoo.com}
should remain active indefinitely, although I don't check it often.

\end{document}